\documentclass[aps, showpacs, showkeys,nofootinbib,floatfix]{revtex4}

\usepackage{amssymb}
\usepackage{amsmath}
\usepackage{graphicx}
\usepackage{hyperref}

\begin{document}
\title{Quasi-Normal Modes of Massless Scalar Field around the 5D Ricci-flat Black String}

\author{Molin Liu}
\email{mlliudl@student.dlut.edu.cn}
\author{Hongya Liu}
\email{hyliu@dlut.edu.cn}
\author{Yuanxing Gui}
\email{guiyx@dlut.edu.cn}

\affiliation{School of Physics and Optoelectronic Technology,
Dalian University of Technology, Dalian, 116024, P. R. China}

\begin{abstract}
As one candidate of the higher dimensional black holes, the 5D
Ricci-flat black string is considered in this paper. By means of a
non-trivial potential $V_{n}$, the quasi-normal modes of a
massless scalar field around this black string space is studied.
By using the classical third order WKB approximation, we analyse
carefully the evolution of frequencies in two aspects, one is the
induced cosmological constant $\Lambda$ and the other is the
quantum number $n$. The massless scalar field decays more slowly
because of the existences of the fifth dimension and the induced
cosmological constant. If extra dimension has in fact existed near
black hole, those quasi-normal frequencies may have some
indication on it.
\end{abstract}



\maketitle

\section{Introduction}
Through an additional field or perturbing the metric itself, the
black hole suffers a damping oscillation phase. Usually, people
call it Quasi-normal models (QNMs) or quasi-normal ringing. So, as
the emission of gravitational wave (GW), a normal model
oscillation is replaced by the complex frequencies
$''\text{quasi-normal}''$ where the real part represents the
actual frequency and the imaginary part represents the damping of
the oscillation. Those frequencies are directly connected to the
black hole's mass, charge, momentum and so on. In the view of
QNMs' evolution, there are three stages: the first one is the
initial outburst from the source of perturbation, the second one
is the damping (quasi-normal) oscillation and the last one is the
asymptotic tails at very late time. The evolution significantly
depends on the asymptotic behavior of the space. In perturbation
theory, the linear perturbation of static black holes was first
studied by Regge and Wheeler in 1957 \cite{Regge}. Soon after
that, Vishveshwara \cite{Vishveshwara} presented the QNMs by
calculating the scattering of gravitational waves (GW) around a
Schwarzschild black hole. Then Press \cite{Press} gave the
original term $quasi-normal$ (QN) $frequencies$. This
perturbations have been studied extensively in many literatures
\cite{QNMs}. For the in-depth reviews, one can refer to
\cite{Nollert} \cite{Kokkotas}. People believe that the QN
frequencies could be detected by GW observatories (LIGO, VIRGO,
TAMT, GEO600, and so on) in the future.

On the other hand, one kind of higher dimensional theory named
induced matter theory is shown by Wesson and co-workers
\cite{Wesson} \cite{Overduin} in the 90's of the last century.
They showed a non-compact fifth dimension and pointed out the 4D
source is induced from an empty 5D manifold. That is, 5D manifold
is Ricci-flat while 4D hypersurface is curved by the 4D matters.
So this theory is also called Space-Time-Matter (STM) theory.
Meanwhile, in the STM framework, there are many extensive
literatures discussing Quantum Dirac Equation \cite{Macias},
Perihelion Problem \cite{Lim}, Kaluza-Klein Solitons
\cite{Billyard} \cite{Liusoliton}, Black Hole \cite{Liu1}
\cite{Liu222} \cite{Mashhoon} \cite{Liu00}, Solar System Tests
\cite{Liu333} and so on.

In the last decade, other robust extra dimensional models have
appeared in gravitational-field theory such as ADD model
\cite{ADD} and Randall-Sundrum I model \cite{Randall2}/ II model
\cite{Randall1} with the additional spacelike dimensions. In those
models, our world is a 3-brane which is embedded in the higher
dimensional space (bulk). To avoid interactions beyond any
acceptable phenomenological limits, standard model (SM) particles
(such as fermions, gauge bosons, Higgs) are confined on a (3 + 1)
dimensional hypersurface (3-brane) without accessing the
transverse dimensions, except for the gravitons and scalar
particles without charges under SM gauge group. In this paper, we
assume the scalar field can freely propagate in the bulk.

If matter trapped on the brane undergoes gravitational collapse, a
black hole will form naturally and its horizon extends into the
extra dimension transverse to brane. Such higher dimensional
object looked like a black hole on the brane is actually a black
string in the higher dimensional brane world \cite{blackstring}.
One natural candidate is a Schwarzschild-de Sitter (SdS) black
hole embedded into the 5D Ricci-flat space \cite{Liu1}
\cite{Mashhoon} \cite{Liu00} \cite{Molin1}. It should be noticed
that the STM theory is equivalent to the brane world \cite{Ponce}
\cite{Seahra} \cite{Liu_plb}.

Meanwhile, one of exciting predictions in large extra dimensional
model \cite{ADD} is that the CERN Large Hadron Collider (LHC) will
produce black holes by the colliding of highly energetic particles
when the scale of quantum gravity is near TeV \cite{LHC}.
Naturally, the detectable QNMs are studied widely in higher
dimensional background \cite{higherQNMs}. In this paper, we
calculate the QN frequency of massless scalar field around a 5D
Ricci-flat black string space.

This paper is organized as follows: In Section II, the 5D
Ricci-flat black string metric and the time-dependent radial
equation about $R_{\omega}(r,t)$ are represented. In section III,
by a tortoise coordinate transformation, the propagating master
equation of scalar field is obtained. In section VI, by using the
third order WKB method, the QN frequencies is obtained in Table I,
II and III. Section V is a conclusion. We adopt the signature $(+,
-, -, -, -)$ and put $\hbar$, $c$ ,and $G$ equal to unity.
Lowercase Greek indices $\mu,\ \nu,\ \ldots$ will be taken to run
over 0, 1, 2, 3 as usual, while capital indices A, B, C, $\ldots$
run over all five coordinates (0, 1, 2, 3, 4).

\section{Klein-Gordon Equation in the 5D Ricci-flat Black String Space}
A class of 5D black holes solutions have been presented by
Mashhoon, Wesson and Liu \cite{Liu1} \cite{Mashhoon} \cite{Wesson}
under STM scenario. Briefly, the static, three-dimensional
spherically symmetric line element takes the form
\begin{equation}
dS^{2}=\frac{\Lambda
\xi^2}{3}\left[f(r)dt^{2}-\frac{1}{f(r)}dr^{2}-r^{2}\left(d\theta^2+\sin^2\theta d\phi^2\right)\right]-d\xi^{2}. \label{eq:5dmetric}%
\end{equation}
where $\xi$ is the open non-compact extra dimension coordinate.
The part of this metric inside the square bracket is exactly the
same line-element as the 4D Schwarzschild-de Sitter solution,
which is bounded by two horizons
--- an inner horizon (black hole horizon) and an outer horizon (one may call this cosmological horizon).

The radial-dependent metric function $f(r)$ takes the form
\begin{equation}
f(r)=1-\frac{2M}{r}-\frac{\Lambda}{3}r^2,\label{f-function}
\end{equation}
where $\Lambda$ is the induced cosmological constant and $M$ is
the central mass. The metric (\ref{eq:5dmetric}) satisfies the 5D
vacuum equation $R_{AB}=0$. Therefore, there is no cosmological
constant when viewed from 5D. But when viewed from 4D, there is an
effective cosmological constant $\Lambda$. So one can treat this
$\Lambda$ as a parameter which comes from the fifth dimension.
This solution has been studied in many works \cite{Mashhoon11}
\cite{Wesson_1} \cite{Liu_2} \cite{Mashhoon_1} focusing mainly on
the induced constant $\Lambda$, the extra force and so on.

We redefine the fifth dimension
$\xi=\sqrt{3/\Lambda}e^{\sqrt{\frac{\Lambda}{3}}y}$. With this
redefinition, the metric (\ref{eq:5dmetric}) can be rewritten as
\begin{equation}
dS^{2}=e^{2\sqrt{\frac{\Lambda}{3}}y}\left[f(r)dt^{2}-\frac{1}{f(r)}dr^{2}-r^{2}\left(d\theta^2+\sin^2\theta d\phi^2\right)-dy^{2}\right],\label{eq:5dmetric-y}%
\end{equation}
Using the line element (\ref{eq:5dmetric}), the metric function
(\ref{f-function}) and above new extra dimension, a
Randall-Sundrum (RS) type brane model is built up. Now, let us
show the configuration in detail. There are two branes in this
model: one brane is at $y=0$ and the other brane is at $y=y_{1}$.
So the fifth dimension becomes finite. It could be very large as
RS II model \cite{Randall2} or very small as RS I model
\cite{Randall1}. The 4D line-element represents exactly the
Schwarzschild-de Sitter black hole on a hypersurface ($\xi$ or $y$
= $constant$). However, viewing from the 5D space, the horizon
does not form a 4D sphere --- it looks like a black string lying
along the extra dimension. So, we call the solution
(\ref{eq:5dmetric}) a 5D Ricci-flat black string solution.

The metric function (\ref{f-function}) can be expressed by the
horizons as follows
\begin{equation}
f(r)=\frac{\Lambda}{3r}(r-r_{e})(r_{c}-r)(r-r_{o}). \label{re-f function}%
\end{equation}

The null hypersurface of this black string space is determined by
its singularity $f(r) = 0$. Obviously, the solutions to this
equation are inner horizon $r_{e}$, outer horizon $r_{c}$ and a
negative solution $r_{o}=-(r_{e}+r_{c})$. The last one has no
physical significance. Here we only consider the real solutions.
$r_{c}$ and $r_{e}$ are given as

\begin{equation}
\left\{
\begin{array}{c}
r_{c} = \frac{2}{\sqrt{\Lambda}}\cos\eta ,\\
r_{e} = \frac{2}{\sqrt{\Lambda}}\cos(120^\circ-\eta),\\
\end{array}
\right.\label{re-rc}
\end{equation}
where $\eta=\frac{1}{3}\arccos(-3M\sqrt{\Lambda})$ with $30^\circ
\leq\eta\leq 60^\circ$. The real physical solutions are accepted
only if
 $\Lambda$ satisfy $\Lambda M^2\leq\frac{1}{9}$ \cite{Liu1}.

The massless scalar field $\Phi$ in the 5D black string space,
satisfies the Klein-Gordon equation $\square\Phi=0$, where $
\square=\frac{1}{\sqrt{g}}\frac{\partial}{\partial
x^{A}}\left(\sqrt{g}g^{AB}\frac{\partial}{\partial{x^{B}}}\right)\label{Dlb}
$  is the 5D d'Alembertian operator. We assume that the separable
solutions are of the form
\begin{equation}
\Phi=\frac{1}{\sqrt{4\pi\omega}}\frac{1}{r}R_{\omega}(r,t)L(y)Y_{lm}(\theta,\phi),\label{wave
function}
\end{equation}
where $R_{\omega}(r,t)$ is the radial time-dependent function,
$Y_{lm}(\theta,\phi)$ is the usual spherical harmonic function.
 The dependent equation about
$R_{\omega}(r,t)$ in QNMs aspect is,
\begin{equation}
 -\frac{1}{f(r)} r^2\frac{\partial^2}{\partial
 t^2}\left(\frac{R_{\omega}}{r}\right)+\frac{\partial}{\partial r}\left(r^2
 f(r)\frac{\partial}{\partial{r}}\left(\frac{R_{\omega}}{r}\right)\right)-\left[\Omega r^2+l(l+1)\right]\frac{R_{\omega}}{r}=0,\label{radius-t-equation}
\end{equation}
where $\Omega$ is the eigenvalue of function $L(y)$. The fifth
dimensional equation about $L(y)$ is
\begin{equation}
    \frac{d^2L(y)}{dy^2}+\Lambda\sqrt{\frac{\Lambda}{3}}\frac{d L(y)}{dy}+\Omega
 L(y)=0,\label{5-th-equation}
\end{equation}
which is discussed carefully in \cite{Liu00}. In the
Randall-Sundrum double branes system, the modes along the extra
dimension are quantized by means of stable standing waves, and
then the eigenvalue is naturally discretized. The discrete spectra
of $L(y)$ is
\begin{equation}\label{Ly}
    L_n (y) = Ce^{-\frac{\sqrt{3\Lambda}}{2}y}\cos
    \left(n\pi\frac{y}{y_1}\right),
\end{equation}
and the quantum parameter $\Omega_{n}$ is
\begin{equation}\label{quan-Omega}
    \Omega_{n}=\frac{n^2\pi^2}{y_{1}^2}+\frac{3}{4}\Lambda,
\end{equation}
where $n=1, 2, 3 \cdots$ and $y_{1}$ is the thickness of the
bulk.
\begin{figure}
  \includegraphics[width=3.5 in]{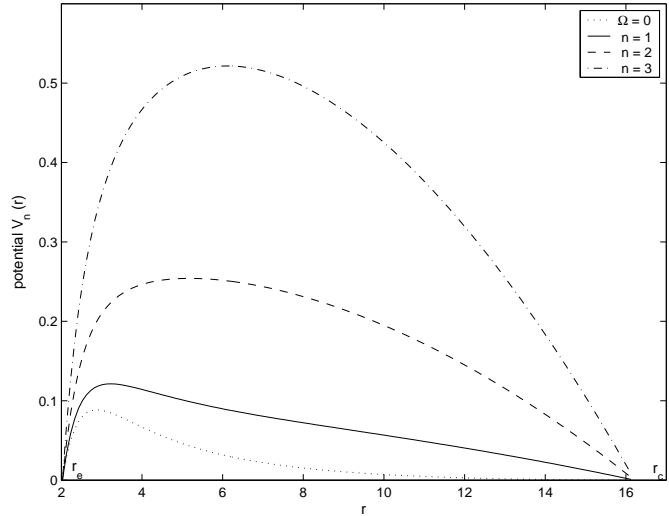}\\
  \caption{The potentials $V_{n}(r)$ versus radial coordinate r with n = 1 (solid), n = 2 (dashed) and n = 3 (dash-dot). Meanwhile, we also draw the case of 4D ($\Omega$ = 0) with the dotted line for
comparison. Here, we adopt $M = 1$, $l = 1$, $\Lambda = 0.01$ and
$y_{1} = 10$ (a very large extra dimension).}\label{potential1}
\end{figure}
\section{The Master Equation for Propagation of Scalar Field in The Bulk}

It is known that radial direction determines the evolution of
black hole radiation. The time variable of Eq.
(\ref{radius-t-equation}) can be removed by the Fourier component
$e^{-i \omega t}$ via
 \begin{equation}
R_{\omega}(r,t)\rightarrow\Psi_{\omega l n}(r) e^{-i\omega t},
 \end{equation}
 where the subscript $n$ presents a new wave function unlike the usual 4D case $\Psi_{\omega l}$ \cite{Brevik}.  Eq. (\ref{radius-t-equation}) can be rewritten as%
 \begin{equation}
 \left[-f(r)\frac{d}{dr}(f(r)\frac{d}{dr})+V(r)\right]\Psi_{\omega
 l n}(r)=\omega^2\Psi_{\omega l n}(r),\label{radius equ. about r}
 \end{equation}
whose potential function is given by%
\begin{equation}
V(r)=f(r)\left[\frac{1}{r}\frac{df(r)}{dr}+\frac{l(l+1)}{r^2}+\Omega\right].\label{potential-of-r}
\end{equation}

\begin{figure}
  \includegraphics[width=3.5 in]{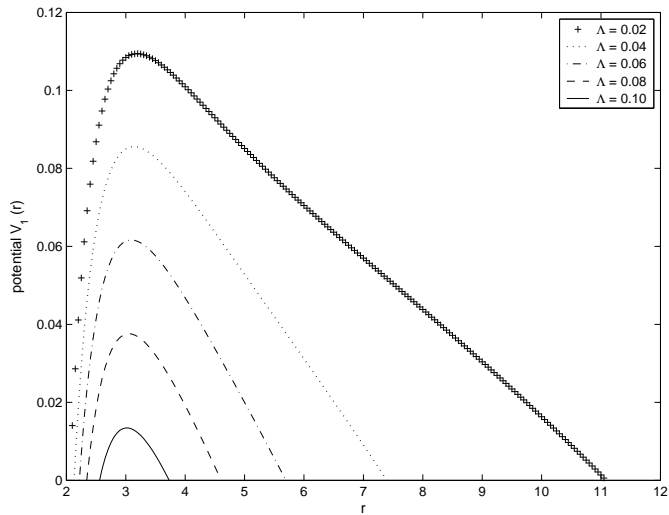}\\
  \caption{The potentials $V_{1}(r)$ versus radial coordinate r with $\Lambda$ = 0.02 (plus), $\Lambda$ = 0.04 (dotted), $\Lambda$ = 0.06 (dash-dot), $\Lambda$ = 0.08 (dashed) and $\Lambda$ = 0.10 (solid). Here, we adopt $M = 1$, $n = 1$, $l = 1$ and
$y_{1} = 10$ (a very large extra dimension).}\label{potential2}
\end{figure}

Now we introduce the tortoise coordinate%
\begin{equation}
x = \int\frac{dr}{f(r)}.\label{tortoise }
\end{equation}
The tortoise coordinate can be expressed with the surface gravity
as follows
\begin{equation}
x=\frac{1}{2M}\left[\frac{1}{2K_{e}}\ln\left(\frac{r}{r_{e}}-1\right)-\frac{1}{2K_{c}}\ln\left(1-\frac{r}{r_{c}}\right)+\frac{1}{2k_{o}}\ln\left(1-\frac{r}{r_{o}}\right)\right],\label{tor-grav-sf}
\end{equation}
where%
\begin{equation}
K_{i}=\frac{1}{2}\left|\frac{df}{dr}\right|_{r=r_i}.
\end{equation}
So under the tortoise coordinate transformation (\ref{tortoise }),
the radial perturbation equation is obtained as%
\begin{equation}
\left[-\frac{d^2}{dx^2} + V(r)\right]\Psi_{\omega l n}(x) =
\omega^2\Psi_{\omega l n}(x).\label{radius-equation}
\end{equation}
It is evident that Eq. (\ref{radius-equation}) is exactly the same
as Ragge-Wheeler equation in QNMs. The incoming or outgoing
particle flowing between inner horizon $r_{e}$ and outer horizon
$r_{c}$ is reflected and transmitted by the potential $V(r)$.
Substituting the quantum parameters $\Omega_{n}$
(\ref{quan-Omega}) into Eq. (\ref{potential-of-r}), the quantum
potentials are obtained as follows
\begin{equation}\label{quan-potential}
    V_{n}(r)=f(r)\left[\frac{1}{r}\frac{df(r)}{dr}+\frac{l(l+1)}{r^2}+\frac{n^2\pi^2}{y_{1}^2}+\frac{3}{4}\Lambda\right],
\end{equation}
which are illustrated in Fig. \ref{potential1} and Fig.
\ref{potential2}. The 5D potential contains the quantum number $n$
which is higher and thicker than the 4D's when $\text{$\Omega$ =
0}$. Also, the height and the thickness of the former increase with
bigger $n$. Meanwhile, for increasing cosmological constant
$\Lambda$, the potential also becomes higher and thicker, and the
interval between $r_{e}$ and $r_{c}$ is larger, too.

According to the quantum potential (\ref{quan-potential}), the
QNMs for massless scalar particles propagating in the black string
space satisfy the boundary conditions \cite{Nollert}
\cite{Kokkotas}
\begin{equation}\label{boundarcondition}
    \Psi_{\omega l n} (x) \approx C_{\pm} \exp(\pm i \omega x) \ \
    \ \ \text{as} \ \ x\longrightarrow\pm\infty,
\end{equation}
denoting pure ingoing waves at the event horizon $r_{e}$ and pure
outgoing waves at cosmological horizon $r_{c}$.
\section{The QN frequency of Massless Scalar Field with the third order WKB Method}
\begin{figure}
  \includegraphics[width=3.5 in]{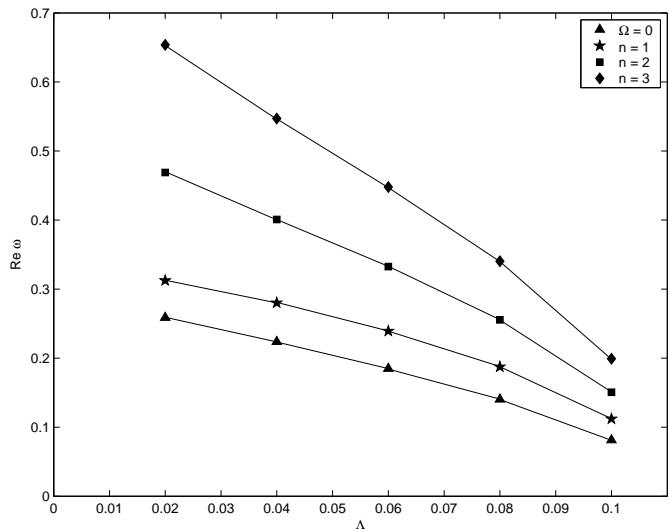}\\
  \caption{The real parts of QN frequencies ($\text{Re }\omega$) of quasi-normal of the scalar field
in 5D Ricci-flat black string space with $l = 1$, $p = 0$, $M = 1$
and $y_{1} = 10$. We denote pentagrams with $n = 1$, squares with
$n = 2$, diamonds with $n = 3$ and triangles with $\Omega =
0$.}\label{fig1}
\end{figure}

\begin{figure}
  \includegraphics[width=3.5 in]{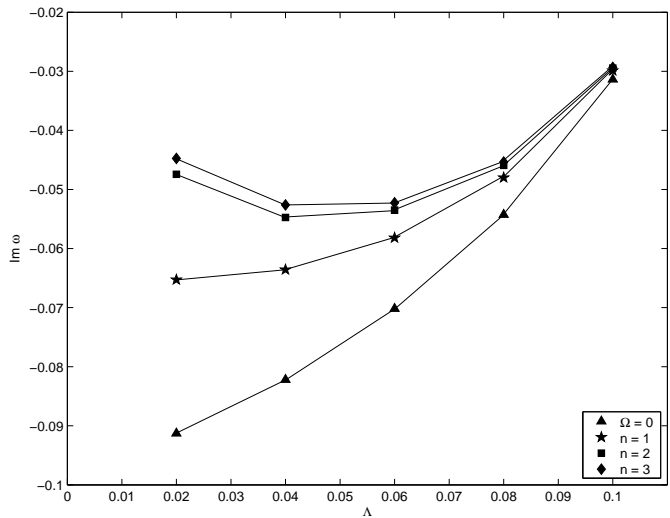}\\
  \caption{The imaginary parts of QN frequencies ($\text{Im }\omega$) of quasi-normal of the scalar field
in 5D Ricci-flat black string space with $l = 1$, $p = 0$, $M = 1$
and $y_{1} = 10$. We denote pentagrams with $n = 1$, squares with
$n = 2$, diamonds with $n = 3$ and triangles with $\Omega =
0$.}\label{fig2}
\end{figure}

Numerical WKB approximation is an effective method to obtain the
complex QN frequencies by using the well-known Bohr-Sommerfeld
rule. It was originally shown by Schutz et al \cite{Schutz} and
was later developed to the third order by Iyer et al \cite{Iyer1}
\cite{Iyer2} and to the sixth order by Konoplya \cite{Konoplya}.
Then after that this method is extensively used in various spaces
\cite{WKB}. The third order WKB formula for QN frequencies has the
form \cite{Iyer1} \cite{Iyer2},
\begin{equation}\label{WKB}
    \omega^2 = \left[V_0 + (-2 V_{0}^{''})^{1/2}
    \tilde{\Lambda}\right] - i \left(p + \frac{1}{2}\right)
    \left(-2 V_{0} ^{''}\right)^{1/2} \left(1 +
    \tilde{\Omega}\right),
\end{equation}

\begin{equation}\label{p}
p = \left\{
\begin{array}{c}
0,\ 1,\ 2,\ \ldots, \text{Re $\omega > 0$},\\
-1,\ -2,\ -3,\ \ldots, \text{Re $\omega < 0$},
\end{array}
\right.
\end{equation}
where
\begin{eqnarray}
  \tilde{\Lambda} &=& \frac{1}{(-2 V_{0}^{''})^{1/2}} \bigg[\frac{1}{8}\bigg[\frac{V_{0}^{(4)}}{V_{0}^{''}}\bigg]\bigg[\frac{1}{4} + \alpha^2\bigg]-\frac{1}{288}\bigg[\frac{V_{0}^{'''}}{V_{0}^{''}}\bigg]^2 (7 + 60\alpha^2)\bigg],\label{lambda}\\
  \nonumber \tilde{\Omega} &=& \frac{1}{(-2 V_{0}^{''})} \bigg[\frac{5}{6912}\bigg[\frac{V_{0}^{'''}}{V_{0}^{''}}\bigg]^4 (77 + 188\alpha^2) - \frac{1}{384} \bigg[\frac{{V_{0}^{'''}}^2 V_{0}^{(4)}}{{V_{0}^{''}}^3}\bigg](51 + 100\alpha^2) + \frac{1}{2304}\bigg[\frac{V_{0}^{(4)}}{V_{0}^{''}}\bigg]^2 (67 +
  68\alpha^2)\\
  & & + \frac{1}{288} \bigg[\frac{V_{0}^{'''} V_{0}^{(5)}}{{V_{0}^{''}}^2}\bigg] (19 + 28\alpha^2) - \frac{1}{288}\bigg[\frac{V_{0}^{(6)}}{V_{0}^{''}}\bigg](5 + 4\alpha^2)
  \bigg],\label{omega}
\end{eqnarray}
where $\alpha\text{ = p + 1/2}$ and symbol $p$ is the various
overtones. The primes and superscript $(n)$ denote differentiation
with respect to the tortoise coordinate $x$. The subscript $0$ on
a variable denotes the value at $x_0$, which is the position of
maximum $V(x)$, namely, $V_{0}^{(n)} = \frac{d^nV}{dx^n}\big|_{x =
x_{0}}$. Substituting potential (\ref{quan-potential}) into
$\tilde{\Lambda}$ and $\tilde{\Omega}$, we can obtain the vital QN
frequencies for the massless scalar field in the 5D black string
space. Meanwhile, it is known that the WKB approximation is
accurate for the low-lying QNM modes, but it fails to calculate
the higher-order modes. Therefore, the condition $l > p$ is
employed and the QN frequencies of fundamental key cases: ($l = 1,
p = 0$), ($l = 2, p = 0$) and ($l = 2, p = 1$) are listed in the
Table I, II and III, respectively. Meanwhile, potential
(\ref{potential-of-r}) illustrate clearly that when $\Omega = 0$
the Regge-Wheeler equation (\ref{radius-equation}) is naturally
reduced to 4D SdS case. So those tables also include a comparison
with the results of 4D case. Here we should notice that the
denotation $\Omega = 0$ does not indicates $n = 0$. To avoid the
confusion about parameters $\Omega$ and $\Omega_n$, we provide
some explanation in the conclusion part. Here, we adopt $M = 1$
and $y_{1} = 10$ and analyse the QN frequencies from two aspects:
cosmological constant $\Lambda$ and quantum number $n$.

\begin{figure}
  \includegraphics[width=3.5 in]{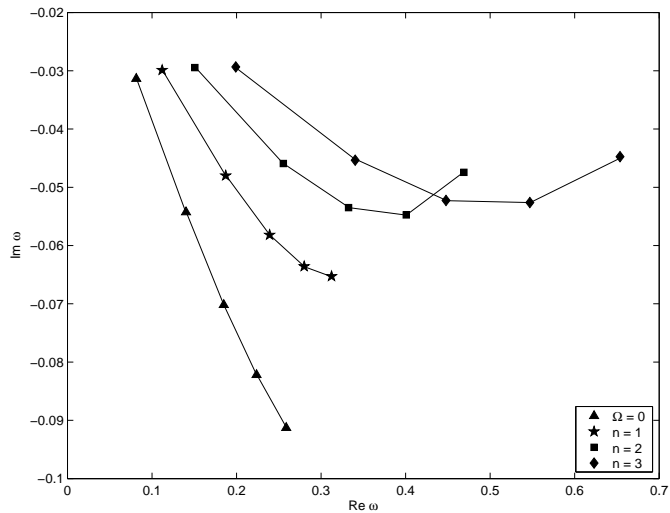}\\
  \caption{The imaginary parts ($\text{Im }\omega$) versus the real
  parts ($\text{Re }\omega$) in 5D Ricci-flat black string space with $l = 1$, $p = 0$, $M = 1$ and
$y_{1} = 10$. We denote pentagrams with $n = 1$, squares with $n =
2$, diamonds with $n = 3$ and triangles with $\Omega =
0$.}\label{fig3}
\end{figure}

\begin{table}[!h]\label{11}
\caption{The QN frequencies of massless scalar field for $l = 1$
and $p = 0$}
\begin{tabular}{|c|c|c|c|c|}
     \hline\hline
           $\Lambda$ & $\omega (n = 1)$ & $\omega (n = 2)$ & $\omega (n = 3)$& $\omega (\Omega = 0)$\\
     \hline
    0.02&$0.312415-0.0652835i$&$0.468963-0.0474421i$&$0.653750-0.0447420i$&$0.258832-0.091277i$\\
     \hline
     0.04&$0.280096-0.0635772i$&$0.400800-0.0547428i$&$0.547038-0.0526103i$&$0.223730-0.082191i$\\
     \hline
     0.06&$0.239281-0.0581832i$&$0.332630-0.0534970i$&$0.447846-0.0522556i$&$0.184973-0.070180i$\\
     \hline
     0.08&$0.187333-0.0479931i$&$0.255402-0.0459195i$&$0.340367-0.0453565i$&$0.140302-0.054267i$\\
     \hline
    0.10&$0.112055-0.0298885i$&$0.150672-0.0294599i$&$0.199165-0.0293426i$&$0.081436-0.031372i$\\
     \hline
     \hline
\end{tabular}
\end{table}

\begin{table}[!h]\label{II}
\caption{The QN frequencies of massless scalar field for $l = 2$
and $p = 0$}
\begin{tabular}{|c|c|c|c|c|}
     \hline\hline
           $\Lambda$ & $\omega (n = 1)$ & $\omega (n = 2)$ & $\omega (n = 3)$& $\omega (\Omega = 0)$\\
     \hline
    0.02&$0.468382-0.0791272i$&$0.561831-0.0581448i$&$0.715167-0.0470975i$&$0.434260-0.0886175i$\\
     \hline
     0.04&$0.415397-0.0714952i$&$0.494992-0.0605678i$&$0.614782-0.0546797i$&$0.380534-0.0787860i$\\
     \hline
     0.06&$0.353220-0.0620269i$&$0.417977-0.0565769i$&$0.512001-0.0535286i$&$0.319866-0.0668696i$\\
     \hline
     0.08&$0.276190-0.0494536i$&$0.324789-0.0472241i$&$0.394093-0.0459458i$&$0.247382-0.0519274i$\\
     \hline
    0.10&$0.165316-0.0301362i$&$0.193354-0.0297108i$&$0.232939-0.0294638i$&$0.146577-0.0306956i$\\
     \hline
     \hline
\end{tabular}
\end{table}

\begin{table}[!h]\label{III}
\caption{The QN frequencies of massless scalar field for $l = 2$
and $p = 1$}
\begin{tabular}{|c|c|c|c|c|}
     \hline\hline
           $\Lambda$ & $\omega (n = 1)$ & $\omega (n = 2)$ & $\omega (n = 3)$& $\omega (\Omega = 0)$\\
     \hline
    0.02&$0.443126-0.242565i$&$0.527013-0.162412i$&$0.718255-0.141176i$&$0.420199-0.268921i$\\
     \hline
     0.04&$0.400020-0.215866i$&$0.482499-0.177642i$&$0.615655-0.163493i$&$0.371026-0.238251i$\\
     \hline
     0.06&$0.345271-0.186311i$&$0.413660-0.168533i$&$0.512501-0.160352i$&$0.313821-0.201780i$\\
     \hline
     0.08&$0.273104-0.148377i$&$0.323641-0.141410i$&$0.394371-0.137765i$&$0.244094-0.156371i$\\
     \hline
    0.10&$0.164801-0.0904138i$&$0.193234-0.0891112i$&$0.233009-0.0883807i$&$0.145667-0.092201i$\\
     \hline
     \hline
\end{tabular}
\end{table}

Firstly, for a given cosmological constant $\Lambda$, it is shown
that the real parts of QNMs ($\text{Re }\omega$) increase with
bigger quantum number $n$. But the absolute value of the imaginary
parts ($|\text{Im }\omega|$) decrease for bigger $n$. In general,
the actual frequencies in 5D are larger than 4D's, and the scalar
field in 5D decays more slowly than the one in 4D. With increasing
$n$, the QN frequencies become larger and the scalar field decays
more slowly.

Secondly, for a given $n$ we can also read that $\text{Re }\omega$
and $|\text{Im }\omega|$ decrease with larger $\Lambda$. It means
that with the increasing cosmological constant $\Lambda$ the
actual frequency becomes smaller and the scalar field decays more
slowly. These results are in agreement with the results of 4D SdS
case with the sixth order WKB method \cite{Zhidenko}. As mentioned
above, the two circumstances for given $n$ and $\Lambda$ can be
manifested in Fig. \ref{fig1} and Fig. \ref{fig2} which are
obtained by Table I.

In order to study the relationship between the actual frequency
and the damping of the oscillation, we also plot the $\text{Im
}\omega$ versus $\text{Re }\omega$ graph in Fig. \ref{fig3}.
Obviously, the absolute value of the imaginary parts increase
entirely with the larger real parts. However, when the quantum
number $n$ becomes larger, there is a break point in the curve. In
other words, $|\text{Im }\omega|$ does not monotonously increase
with $\text{Re }\omega$ for bigger $n$, especially in the case of
$n =2,\ 3$.

To explain the reason, we should refer to this black string's
reflection (or transmission) \cite{Molin1}. In the square barrier
model \cite{Molin1}, the reflection or transmission coefficients
have the analytic forms. As one knows that the reflection should
be stronger with higher barriers. The reflection coefficients of
$n=2,\ 3$ would be larger than the cases of $n=1$ or $\Omega = 0$
in the usual viewpoint. On the contrary, the values of reflection
coefficients of $n = 2,\ 3$ are smaller than the cases of $n=1$ or
$\Omega = 0$. If the resonant effect of quantum mechanics exists
in the barriers, the peculiar behavior is easily to be gotten.
Mathematically, an oscillatory cosine function $\cos(2k2d)$ is
contained in the expression of reflection coefficients $R$ or
transmission coefficients $T$ \cite{Molin1}. (For detailed
discussion of those behavior, see Ref. \cite{Molin1}.) The
scatting potential of $n = 2,\ 3$ also have the peculiar QN
frequencies by the same resonant effect. From Fig.
\ref{potential1}, we can read that with larger $n$ the peak of
potential slips the cosmological horizon and the potential becomes
higher and wider, especially in the case of $n =2,\ 3$. Hence with
the enhancing of resonant effect the QN frequencies decay more
slowly, even though the QN frequencies become larger. So the
anomalous values of $n=2,\ 3$ are the result of resonant effect of
quantum mechanics in the barriers.

\section{conclusion }
In this paper, we have used the third-order WKB approximation to
calculate the quasi-normal frequencies of massless scalar field
outside a 5D black string. We summarize what has been achieved.

1. In this 5D Ricci-flat black string space, the QNMs is studied
by fixing either the cosmological constant $\Lambda$ or the
quantum number $n$. From the result we find that the scalar field
decays more slowly with the increasing $\Lambda$ or $n$. For a
given cosmological constant $\Lambda$, the 5D actual frequency
becomes bigger with increasing $n$. While for a given $n$, the
frequency becomes smaller with bigger cosmological constant
$\Lambda$. In other words, the 5D QN frequencies are larger than
4D's ($\Omega = 0$).

2. As one candidate of the higher dimensional black hole, the 5D
Ricci-flat black string implies us something interesting. The
quantum number $n$ depicts a new wave solution $\Psi_{\omega l n}$
of Schr$\ddot{o}$dinger wavelike equation. The spectrum of
original potential is discrete for the existence of quantum number
$n$. The non-trivial radiation can reveal much of valuable
information that characterizes the higher dimensional background,
such as the dimensionality of space, the topological structure and
so on. Here, the QN frequencies of 5D Ricci-flat black string are
determined by the black hole mass $M$, the effective cosmological
constant $\Lambda$, the quantum number $n$ and the thickness of
the bulk $y_{1}$. The information about extra dimension is encoded
in those QN frequencies such as the magnitude of extra dimension
and the thickness of the bulk. It is known that the best method to
probe black hole is the detectable QN spectrum. If extra dimension
does exist and is visible near black hole, maybe those QN
frequency can prove its existence.

3. To ensure the validity of results, tables and figures, we use
Mathematica software to design a program and calculate carefully
those QN frequencies. The induced four dimensional results
($\Omega = 0$) are exactly identical with 4D SdS black hole's
\cite{Zhidenko}. Otherwise, just as 4D case \cite{Zhidenko}, the
QN frequencies decrease with increasing cosmological constant
$\Lambda$. Of course, this program is tested in some other black
holes such as 4D Schwarzschild black hole \cite{Iyer2}, and the
same results are got which is not presented in this paper. This
method gets desired effects and hence believable. .

4. The reason why we discuss, not the gravitational perturbation
but a test scalar field, is that this paper is a continuation of
previous work \cite{Liu00} to a certain extent. As we known from
the spirit of string theory, the standard model fields are
confined on 3-brane except for gravitons and scalar particles. The
original goal to introduce the scalar field is to examine the
effect of an extra dimension on black hole radiation.
Surprisingly, the radial component of 5D Klein-Gordon equation can
be rewritten exactly as the Regge-Wheeler form. Considering the
QNMs boundary condition, we have studied the spectrum of QN
frequencies by usual third order WKB method. After all, the QNMs
of higher dimensional black hole is very attractive and
interesting. Certainly, the basic gravitation field is easier to
be detected by gravitational wave than other fields. Anyway, it is
interesting to study this case and further work is needed.

5. It should be noticed that the parameter $\Omega$ is not the
same as $\Omega_n$. In fact, parameter $\Omega$ is introduced to
separate the variables $R_{\omega}(r,t)$, $L(y)$ and
$Y_{lm}(\theta,\phi)$ in this paper. But parameter $\Omega_n$ is a
particular eigenvalue which is deduced from the original $\Omega$
under the standing wave condition \cite{Liu00},
\begin{equation}\label{add1}
y_1 \sqrt{\Omega - \frac{3}{4}\Lambda} = n\pi.
\end{equation}
In other words, $\Omega$ is a free parameter, while $\Omega_n =
\frac{n^2\pi^2}{y_1^2} + \frac{3}{4}\Lambda$ is constrained by
$n$, $y_1$ and $\Lambda$. Meanwhile, the quantum number $n$ is a
positive integer i.e. $n > 0$ according to the condition
(\ref{add1}) \cite{Liu00}. Furthermore, cosmological constant
$\Lambda$ is nonzero in the de Sitter universe. So considering the
conditions $n > 0$ and $\Lambda > 0$, we can get the eigenvalue
$\Omega_n > 0$. By the way, when $n = 0$ the steady stand wave
(\ref{Ly}) can not be formed and this case must be abandoned. On
the other hand, the potential function (\ref{potential-of-r})
indicates that the case $\Omega = 0$ (not n = 0) corresponds to
usual 4D SdS. Hence, the notations $\Omega = 0$ in figures and
tables just show the one of 4D SdS.  Of course, some other
notation could be used in principle.

 \acknowledgments Project supported by the National Basic
Research Program of China (2003CB716300), National Natural Science
Foundation of China (10573003) and National Natural Science
Foundation of China (10573004). We are grateful to Feng Luo for
useful help.

\end{document}